# Projections of Earth's Technosphere: Strategies for Observing Technosignatures on Terrestrial Exoplanets

Jacob Haqq-Misra[1], Ravi K. Kopparapu[2], and George Profitiliotis[1]
[1] Blue Marble Space Institute of Science, Seattle, WA 98104, USA; jacob@bmsis.org
[2] NASA Goddard Space Flight Center, Greenbelt, MD 20771, USA

## Abstract

The search for technosignatures—remotely detectable evidence of extraterrestrial technology—draws upon examples from the recent history of Earth as well as projections of Earth's technosphere. Facilities like the Habitable Worlds Observatory (HWO) will significantly advance the feasibility of characterizing the atmospheres of habitable exoplanets at visible and near-infrared wavelengths, while other future mission concepts could extend this search to mid-infrared wavelengths. We draw upon a recently developed set of 10 self-consistent scenarios for future Earth technospheres as analogs for extraterrestrial technospheres, which we use to outline a stepwise technosignature search strategy, beginning with HWO and followed by other missions. We find that HWO could reveal elevated abundances of a $CO_2$ + $NO_2$ pair on planets with combustion and other large-scale industry, which could be observable in up to eight of the 10 scenarios. Follow-up radio observations could reveal narrowband directed transmissions, as occur in two of the scenarios. Further study involving direct detections at mid-infrared wavelengths with the Large Interferometer for Exoplanets could reveal spectral features from industry, such as the combinations of $CO_2$ + CFC-11/12 in four scenarios and $CO_2$ + CFC-11/12 + $CF_4$ in one scenario; two of these also include the $N_2O$ + $CH_4$ + $NH_3$ triple from large-scale agriculture. Other mission concepts, such as a solar-gravitational-lens mission, could reveal large-scale surface features in two scenarios that would otherwise show no detectable technosignatures, while an interplanetary flyby or probe mission would be the most conclusive way to reveal the presence of technology on terrestrial exoplanets.

*Unified Astronomy Thesaurus concepts:* Technosignatures (2128); Astrobiology (74); Search for extraterrestrial intelligence (2127); Exoplanets (498); Habitable planets (695)

## 1. Introduction

Earth is the only known example of a planet with life, as well as the only planet with global-scale technological signatures. Therefore, the search for extraterrestrial life must begin by considering the possible biosignatures and technosignatures that have existed in Earth's history or may exist in Earth's future. The reliance on terrestrial examples does not imply that any extraterrestrial biospheres or technospheres will necessarily resemble those of Earth; instead, these examples of known biology and technology provide a basis for thinking about the technical requirements that would be needed to detect such signatures in exoplanetary systems. In other words, the example of Earth as an inhabited planet can be used for defining search strategies, designing ground- or space-based telescopes, and motivating theoretical work that would enable us to recognize an Earth-like biosphere or technosphere using the tools of astronomy. Such tools may also be relevant for detecting features of exotic and non-Earth-like planets, possibly including biospheres and technospheres that bear no resemblance to Earth; however, because such possibilities are as yet unknown, they cannot be of direct use in the design of instrumentation or search strategies. For better or worse, Earth itself remains the primary point of reference for thinking about a systematic approach toward looking for extraterrestrial life.

Earth has been inhabited for most of its geologic history, so the search for extraterrestrial biospheres can draw upon Earth today (e.g., T. D. Robinson et al. 2011; M. F. Sterzik et al. 2012; J. Lustig-Yaeger et al. 2023) as well as Earth through time (e.g., G. Arney et al. 2016, 2018; E. W. Schwieterman et al. 2018; L. Kaltenegger et al. 2020; A. Tokadjian et al. 2024), to provide examples of atmospheric spectral signatures that could reveal the presence of life. Theoretical studies have also attempted to expand beyond these known examples to consider other possibilities for spectral features that could be detected from extraterrestrial biospheres that are much less like Earth (J. Krissansen-Totton et al. 2016, 2018) as well as potential false-positive spectral biosignatures (D. C. Catling et al. 2018; C. E. Harman & S. Domagal-Goldman 2018; V. S. Meadows et al. 2018; S. Foote et al. 2023). These efforts all seek to draw upon knowledge of Earth's biosphere from the Archean to the present as a way of generating a possibility space for biosignatures that could be remotely detectable in exoplanetary systems. The actual detection of an extraterrestrial biosphere may not necessarily bear any strong resemblance to such historical or theoretical possibilities, but these examples nevertheless provide ideas about how to identify the presence of life through the atmospheric spectral features of an exoplanet.

Technology is a relatively young phenomenon on Earth, so the search for extraterrestrial technospheres can only draw upon Earth's very recent history to provide examples of spectral signatures that could reveal the presence of technology. Possible technosignatures on Earth today that could conceivably be remotely detectable include the spectral absorption features of atmospheric pollution from nitrogen dioxide (R. Kopparapu et al. 2021), chlorofluorocarbons (J. Haqq-Misra et al. 2022b; J. Lustig-Yaeger et al. 2023), and large-scale agriculture (J. Haqq-Misra et al. 2022a), as well as the emission spectrum of nightside artificial illumination (A. Loeb & E. L. Turner 2012; T. G. Beatty 2022). Other theoretical studies have extended this possibility space by considering potential future developments that could arise on Earth and lead to detectable technosignatures, which include the reflectance spectrum from the large-scale deployment of photovoltaics (M. Lingam & A. Loeb 2017; R. Kopparapu et al. 2024), mid-infrared absorption features of artificial greenhouse gases used for terraforming (M. Elowitz 2022; J. Haqq-Misra et al. 2022b; S. Seager et al. 2023; E. W. Schwieterman et al. 2024), ultraviolet absorption features that arise from elevated abundances of atmospheric polycyclic aromatic hydrocarbons (D. Dubey et al. 2025), long-lived atmospheric radionuclides that persist from nuclear activity (A. Stevens et al. 2016; J. Haqq-Misra et al. 2024), and evidence of a depleted D/H ratio in atmospheric water vapor indicative of fusion technology (D. C. Catling et al. 2025). Even more speculative suggestions include observations of anomalous infrared excesses that could be evidence of megastructures, such as Dyson spheres (F. J. Dyson 1960; J. T. Wright 2020) that are designed to collect much a larger fraction of emitted starlight than would be incident on the planet alone. Any actual extraterrestrial technology that does exist may not necessarily resemble any of these possibilities, but these examples drawn from technology on Earth still suggest observation strategies that could be applied to the search for technosignatures on exoplanets (see, e.g., H. Socas-Navarro et al. 2021; J. Haqq-Misra et al. 2022c; M. Lingam et al. 2023; S. Z. Sheikh et al. 2025).

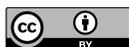

Many technosignature studies attempt to extrapolate current trends on Earth as a way of exploring scenarios in which technology might be more easily detectable than on present-day Earth. Examples include detectability calculations for elevated levels of atmospheric pollutants (e.g., R. Kopparapu et al. 2021; J. Haqq-Misra et al. 2022a, 2022b; D. Dubey et al. 2025) or expansive areas of artificial illumination (e.g., T. G. Beatty 2022) on exoplanets that are orders of magnitude above those on Earth today. The logic in such an approach is that technology on Earth represents examples of phenomena that are known, so it might be reasonable to imagine continued trajectories of current trends that could lead to such outcomes. This informal approach has been routinely invoked since the origin of the Search for Extraterrestrial Intelligence (SETI) in the 1960s; however, such assumptions tend to neglect the couplings that exist between human systems and technology, focusing instead on linear or exponential extrapolations in trends, like the rates of growth or consumption, which may or may not be tenable.

The interdisciplinary field of futures studies provides a more robust set of tools for developing self-consistent approaches for thinking about future trajectories. The field uses the word "futures" in the plural sense to indicate that it is an approach for making *projections* of several possible future scenarios that are tenable rather than an attempt to *predict* the actual future. Futures studies methods are widely used in applications including climate change, strategic defense, urban planning, and product development, among others, which usually make projections on timescales of years to decades. In the search for technosignatures, futures studies methods can enable more systematic thinking about the range of possible technospheres that could arise in Earth's future, taking into account the interdependencies of human and technological systems and the deep uncertainty involved in how such systems can develop (J. Voros 2018). With this goal in mind, the study by J. Haqq-Misra et al. (2025a) leveraged existing methods from futures studies to develop a set of 10 self-consistent scenarios for Earth's technosphere 1000 yr from now. These 10 scenarios all contain detailed descriptions of unique technospheres that have a logical continuation from Earth today and also connect any features of the future technosphere to basic human needs. The methodological approach is described in detail by J. Haqq-Misra et al. (2025a), including an overview of each scenario and a description of the method for ascertaining the properties of the technosphere. For the purposes of this paper, these 10 scenarios provide examples of the possibility space that could exist for Earth's future technosphere, which can also serve as analogs for thinking about the detectability of extraterrestrial technospheres.

This paper outlines a strategy for observing technospheres on terrestrial exoplanets by assessing the detectability of technosignatures in each of the 10 scenarios developed by J. Haqq-Misra et al. (2025a). This observation strategy will focus primarily on the use of the Habitable Worlds Observatory (HWO), which will be capable of directly imaging Earth-sized planets orbiting Sun-like stars (FGK stellar spectral types) in reflected light, but it will also consider the relevance of follow-up observations by radio telescopes (e.g., G. Harp et al. 2016; N. Franz et al. 2022), as well as the capabilities of future tools, such as the Large Interferometer for Exoplanets (LIFE; S. P. Quanz et al. 2022) mission concept and the solar gravitational lens (S. G. Turyshev & V. T. Toth 2022). The relevance of this approach can be considered in two ways. The first way is to imagine that each of these 10 scenarios represent actual discoveries of habitable exoplanetary systems, with architectures similar to the solar system. In this case, the following discussion demonstrates the ability (or inability) of upcoming missions to reveal the presence of technospheres in each scenario. The second way is to imagine that a single habitable exoplanetary system has been discovered and flagged for intensive follow-up observations. In this case, the following discussion demonstrates how the set of 10 scenarios can serve as templates for examining the possibility of technosignatures in this system. In either case, the purpose of this paper is to assess the capability of future observatories to detect future-Earth-analog technosignatures, although some systems with expansive technospheres may still remain (in the words of author Karl Schroeder) "indistinguishable from nature."

## 2. Characterization by HWO

HWO is NASA's next flagship mission that was recommended by the 2020 Decadal Survey on Astronomy and Astrophysics, with a target launch date in the 2040s. The recommendation from the survey committee was a large-aperture space telescope with a ∼6–8 m mirror that spans ultraviolet, optical, and near-infrared wavelengths, ranging from about 0.2 to 1.8 $\mu$m, capable of performing transformative astrophysics. HWO will be the first telescope designed with optimized capabilities to search for signs of life on planets outside the solar system, specifically by performing direct imaging of exoplanetary systems in reflected light. The primary objective of HWO will be to identify and directly image at least 25 potentially habitable terrestrial planets orbiting nearby F, G, and K dwarf stars, which could reveal the presence of biosignature candidates on some planets or else place statistical limits on the presence of life in nearby systems (D. Angerhausen et al. 2025).

HWO will provide the first opportunity to search for biosignatures and technosignatures in the atmospheres of Earth-sized planets orbiting within the liquid water habitable zone of G dwarf host stars. Current efforts to characterize the atmospheres of habitable terrestrial planets are limited to transiting M dwarf systems. For observatories like the James Webb Space Telescope (JWST) or upcoming extremely large telescopes, any terrestrial planets that orbit the habitable zones of G dwarf stars are beyond the inner working angle for direct imaging; likewise, the transit depth and probability are too small for such planets. Atmospheric characterization of transiting G dwarf planets by current missions may only be possible for hot planets, like 55 Cancri e, which are unlikely to be habitable. As a result, characterization efforts by JWST of potentially habitable exoplanets have focused on transiting systems such as TRAPPIST-1 (e.g., A. P. Lincowski et al. 2023; J. de Wit et al. 2024; E. Ducrot et al. 2025), and near-term efforts will continue to focus on nearby transiting M dwarf systems. M dwarf candidates cannot yet be ruled out as possible sites for harboring life (e.g., A. L. Shields et al. 2016; A. Wandel 2018; H. Chen et al. 2019; N. Madhusudhan et al. 2021) or technology (e.g., J. Haqq-Misra & T. J. Fauchez 2022), but such systems are considerably different from Earth. Given that Earth is the only known example of an inhabited planet, the search for extraterrestrial life must necessarily include a comprehensive search of habitable planets orbiting G dwarf host stars in order to sufficiently examine the hypothesis that biogenesis and technogenesis have occurred elsewhere.

The set of 10 future-Earth scenarios developed by J. Haqq-Misra et al. (2025a) all assume a planet orbiting a G dwarf host star with a solar system architecture. The use of these scenarios as analogs for examining the detectability of extraterrestrial technospheres does not necessarily imply that such an architecture is likely for hosting life or technology. Instead, the scenario-building process utilized by J. Haqq-Misra et al. (2025a) began with present-day Earth, to make projections about plausible and self-consistent future trajectories of human civilization, which assume that humans have remained within the solar system and have not gone extinct. The methodology for developing these scenarios is predicated upon the basic human needs that emerge in each scenario, depending on a planetary system environment that includes a G dwarf host star. It may be beneficial to consider extensions of such futures studies methods for hypothetical Earth-originating civilizations that have migrated from the Sun to an M dwarf host star or a different planetary architecture. The results of this present study could be used as a starting point for such an activity, but further exploration will be reserved for later work.

This section considers the potential for observing technosignatures in these 10 scenarios of Earth's future with the capabilities of HWO. Each scenario includes a description of the atmospheric composition and other features of the technosphere, which can be used to generate synthetic reflected-light spectra for each case and estimate the spectral technosignature features that could be observable by HWO.

The spectral signatures of each scenario will primarily be considered for Earth, given that Earth orbits within the habitable zone of the Sun and that all 10 scenarios begin with Earth's present-day technosphere. HWO may also be able to provide reflected-light spectra for Venus in a system with a solar system architecture, and in some scenarios imaging these other planets could be needed to reveal evidence of extraterrestrial technology.

### 2.1. Observations of Earth

Each scenario developed by J. Haqq-Misra et al. (2025a) includes specifications about the atmospheric composition of Earth, including abundances of greenhouse gases linked to technological activity as well as other elements of the technosphere that could be remotely detectable. All of these features were developed using a self-consistent methodology, in which the features of the technosphere are derived from basic human needs in each scenario and exist for purposes based on the specific details of each scenario. The narrative details of each scenario are omitted here but are described thoroughly by J. Haqq-Misra et al. (2025a; see the main text and supplementary information).[3] For the purposes of this paper, the relevant atmospheric properties and other technosignatures for each scenario relevant to HWO are summarized in Table 1. Each scenario is assigned a numbered identifier, ranging from S1 to S10, which corresponds to the identifiers used by J. Haqq-Misra et al. (2025a). These properties of Earth's atmosphere and technosphere provide a basis for assessing potential detectability with HWO.

The atmospheric constituents listed in Table 1 are taken as steady-state abundances for future Earth in each scenario. In principle, the best approach for considering such atmospheres would be to perform calculations using a coupled climate–chemistry model, in order to generate vertical profiles and account for interactions between chemical species. However, few (if any) such models exist that can account all of the industrial greenhouse gas species in Table 1 (i.e., CFC-11, CFC-12, $CF_4$, $SF_6$, and $NF_3$); likewise, the use of such a model may only provide incremental improvements to our assessment of detectability. We therefore save such calculations for future work and instead make several simplifying assumptions about the atmospheres of Earth in these 10 scenarios. We calculate the global mean temperature, $T$, for Earth in each scenario by assuming that carbon dioxide is the primary greenhouse gas, so that $T = 288 \text{ K} + \Delta T_{CO_2}$, where 288 K is the preindustrial global mean temperature and $\Delta T_{CO_2}$ is the temperature change due to the accumulation of carbon dioxide above preindustrial levels. The value of $\Delta T_{CO_2}$ is calculated following a simplified expression that was developed by G. Myhre et al. (1998):

$$\Delta T_{CO_2} = \lambda \Delta F_{CO_2} = \lambda a \ln \frac{fCO_2}{fCO_2^0}, \quad (1)$$

where $\lambda \approx 1 \text{ K } (\text{W}^{-1} \text{m}^{-2})$ is the climate sensitivity, $fCO_2^0 = 280$ ppmv is the preindustrial carbon dioxide mixing ratio, and $a = 5.35 \text{ W m}^{-2}$. Equation (1) remains accurate to within 1% of more recent revisions of this expression for historical forcings and moderate values of $fCO_2$ (M. Etminan et al. 2016). We otherwise assume Earth-like conditions for the

---
[3] See also futures.bmsis.org for further details.

**Table 1**
Atmospheric Properties for Future Earth in Each of the 10 Scenarios (S1–S10), with Reference Values for Present-day Earth (R1) and Preagricultural Earth (R0)

| | R0 | R1 | S1 | S2 | S3 | S4 | S5 | S6 | S7 | S8 | S9 | S10 |
|---|---|---|---|---|---|---|---|---|---|---|---|---|
| Mean Temperature (K) | 288 | 290 | 300 | 300 | 293 | 288 | 288 | 300 | 289 | 296 | 288 | 288 |
| $CO_2$ (ppm) | 280 | 420 | 11,000 | 3200 | 750 | 290 | 280 | 30,000 | 350 | 1400 | 280 | 280 |
| $CH_4$ (ppm) | 0.57 | 1.9 | 0.57 | 21 | 2.2 | 0.96 | 0.57 | 0.11 | 1.3 | 4.6 | 0.57 | 0.57 |
| $NO_x$ (ppb) | 0.1 | 2 | 150 | 40 | 6.6 | 0.17 | 0.1 | 400 | 1.1 | 15 | 0.1 | 0.1 |
| $N_2O$ (ppb) | 170 | 340 | 170 | 2700 | 370 | 220 | 170 | 34 | 260 | 680 | 170 | 170 |
| $NH_3$ (ppb) | 2 | 10 | 2 | 120 | 12 | 4.3 | 2 | 0.4 | 6.2 | 26 | 2 | 2 |
| CFC-11 (ppb) | ⋯ | 0.23 | 18 | 4.8 | 0.78 | ⋯ | ⋯ | 48 | ⋯ | 1.8 | ⋯ | ⋯ |
| CFC-12 (ppb) | ⋯ | 0.52 | 41 | 11 | 1.8 | ⋯ | ⋯ | 110 | ⋯ | 4.1 | ⋯ | ⋯ |
| $NF_3$ (ppt) | ⋯ | 2.5 | 200 | 53 | 8.5 | 2.5 | ⋯ | 530 | 2.5 | 19.5 | ⋯ | ⋯ |
| $SF_6$ (ppt) | 0.01 | 11 | 870 | 230 | 37 | 11 | 0.01 | 2300 | 11 | 86 | 0.01 | 0.01 |
| $CF_4$ (ppt) | 35 | 87 | 4100 | 1100 | 210 | 87 | 35 | 11,000 | 87 | 440 | 35 | 35 |
| $SO_2$ (ppb), Stratospheric | ⋯ | ⋯ | 16 | 4 | ⋯ | ⋯ | ⋯ | 26 | ⋯ | ⋯ | ⋯ | ⋯ |
| Na Emission (erg s$^{-1}$ cm$^{-2}$) | ⋯ | 0.071 | 18.2 | 2.55 | 0.67 | 0.042 | 56.5 | 0.050 | ⋯ | ⋯ | ⋯ | ⋯ |
| Laser Emission, 1.064 $\mu$m (MW) | ⋯ | ⋯ | 10 | ⋯ | ⋯ | ⋯ | ⋯ | ⋯ | ⋯ | ⋯ | ⋯ | ⋯ |

**Note.** The mixing ratios for greenhouse gases were tabulated by J. Haqq-Misra et al. (2025a). The addition of the $SO_2$ stratospheric aerosol is included in three scenarios (S1, S2, and S6) as solar radiation modification, to compensate for excessive greenhouse warming. Seven scenarios (S1, S2, S3, S4, S6, S7, and S8) include atmospheric pollutants from industry. Six scenarios (S1, S2, S3, S4, S5, and S6) include emission from high-pressure sodium lights in urban areas, and one scenario (S1) includes directed laser emissions. Two scenarios (S9 and S10) have the same properties as preagricultural Earth. The global mean temperature is calculated by assuming that $CO_2$ is the dominant greenhouse gas (Equation (1)). All scenarios otherwise assume a background pressure of 1 bar composed of 78% $N_2$, 21% $O_2$, and 1% Ar.

model atmosphere and neglect any changes that would affect the average water vapor content or otherwise affect the vertical atmospheric structure.

The set of artificial greenhouse gases listed in Table 1 is intended to serve as a representation of long-lived pollutants that are known to occur from industrial processes and that can exert greenhouse warming; although other industrial species are possible, even on Earth today, the specific set of CFC-11, CFC-12, $CF_4$, $SF_6$, and $NF_3$ exemplifies the effects on detectability that can arise from the accumulation of such pollutants. Prior studies (e.g., M. Elowitz 2022; J. Haqq-Misra et al. 2022b; S. Seager et al. 2023; E. W. Schwieterman et al. 2024) have noted that the spectral features of these industrial greenhouse gases are primarily at mid-infrared wavelengths, beyond the observable range of HWO.

Three scenarios (S1, S2, and S6) include solar radiation modification to offset the greenhouse warming from the excessive accumulation of carbon dioxide and other greenhouse gases. Solar radiation modification in these scenarios is accounted for by the injection of sulfur dioxide aerosol into the stratosphere, with a simplifying assumption that 10 Tg of $SO_2$ corresponds to a reduction of 1 K in the global mean temperature (see D. MacMartin et al. 2022). Solar radiation modification is needed in these scenarios in order to avoid triggering a runaway greenhouse state, which could occur at $CO_2$ abundances ∼12 times preindustrial values (R. M. Ramirez et al. 2014). The stratospheric $SO_2$ abundances in these three scenarios correspond to the values required to reduce the global mean temperature to levels that would occur from about 8 times the preindustrial values of $CO_2$.

Many of the scenarios include artificial illumination, which is represented as emission from high-pressure sodium lights, with a value of 0.071 erg s$^{-1}$ cm$^{-2}$ for present-day Earth and 62.8 erg s$^{-1}$ cm$^{-2}$ for a planet-spanning "ecumenopolis" that covers both ocean and land (T. G. Beatty 2022). The values for artificial illumination in Table 1 also include adjustments for certain scenarios (S6, S7, and S8) in which urban lighting has been intentionally directed downward or dimmed to minimize light pollution.

One scenario (S1) includes a continuously broadcast optical signal toward nearby habitable planets, specifically for the purpose of establishing contact with other life (i.e, a laser pulse for messaging to extraterrestrial intelligence or METI). These optical transmissions are assumed to occur at 1.064 $\mu$m (see M. Hippke 2018; S.-Y. Narusawa et al. 2018) at a power of 10 MW. Several scenarios also include optical communication between planets and from planets to satellites (S1, S2, S3, S5, S6, S8, S9, and S10), and one scenario includes the use of laser-propelled nanocraft; however, much work needs to be done to estimate if HWO would be able to see these features in the observed spectra. As this requires some extensive and careful analysis, we leave it for future work.

Synthetic spectra for future Earth in each scenario as they would be observed by HWO are calculated using the Planetary Spectrum Generator (PSG; G. L. Villanueva et al. 2018, 2022) and shown in Figures 1 and 2. Calculations are also performed for present-day Earth (R1) and preagricultural Earth (R0) for comparison (Figure 1). These simulations assume the target is an Earth-like planet 10 pc away. In particular, the most obvious difference between the spectra is the magnitude of $NO_2$ production arising from combustion (as well as large-scale agriculture), which is absent from preagricultural Earth. Whether or not this level of $NO_2$ on the present-day Earth is detectable with HWO-like telescopes remains to be calculated. Although R. Kopparapu et al. (2021) indicated that a 15 m LUVOIR-like telescope (a previous concept precursor study to HWO) could detect current Earth $NO_2$ levels at a signal-to-noise ratio of 5 in ∼400 hr of observation time, more accurate detectability estimates taking into account the current recommended mirror size of HWO (6 m), along with appropriate retrieval analyses, are crucial (but beyond the scope of this study). If it turns out that HWO is indeed able to detect current-Earth-level $NO_2$, this would be a significant result, because HWO would then represent humanity's first

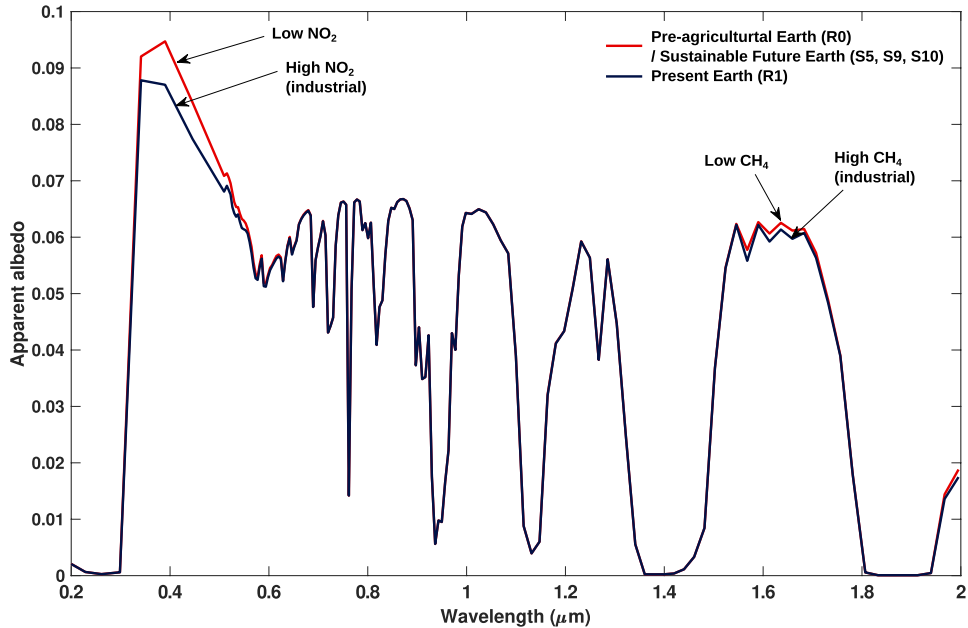

**Figure 1.** Albedo spectra of preagricultural Earth (red line; scenario R0 from Table 1) and present-day Earth (dark line; scenario R1 from Table 1) observed 10 pc away at nominal HWO wavelengths. The noticeable difference between the spectra in the 0.3–0.6 $\mu$m region arises from the significant production of $NO_2$ pollution from industrial activity, as well as large-scale agriculture, on present-day Earth. Note that we assume an isoprofile abundance for the $NO_2$ gas concentration, which overestimates the absorption feature for the present-day Earth. The small difference between the spectra near 1.6 $\mu$m is due to the excess production of $CH_4$ from human activity (including agriculture, combustion, and decomposition of landfill waste). Scenarios S5, S9, and S10 are spectrally identical to preagricultural Earth (R0).

attempt to simultaneously search for both biosignatures and technosignatures.

Additional scenarios from Table 1 are shown in Figure 2, compared to the present-day-Earth case (R1). Most of them have distinct features that are noticeable in the visible and near-infrared parts of the spectrum, which arise from industrial by-products of combustion, such as $NO_2$ and $CO_2$. These spectral features from industrial activity in scenarios S1, S2, S3, S4, S6, S7, and S8 could potentially be observable by HWO. We have not plotted scenarios S5, S9, and S10, as they are identical at this wavelength range to the preagricultural scenario (R0) shown in Figure 1. The combination of $NO_2$ and $CO_2$ absorption, particularly for the strongest cases (S1, S2, and S6), would be evidence of a technosignature; however, further work is needed to explore possible false-positive scenarios that could lead to similar spectral features through nontechnological processes. Likewise, a thorough analysis of retrieval studies is needed to definitively assess the detectability of the features, and we leave that for future work.

The values in Table 1 indicate that both S1 and S5 show strong sodium emission lines that result from artificial illumination, due to widespread urbanization, while S1 also includes emission at 1.064 $\mu$m from communicative laser pulses. These features, if observed by HWO, would appear as anomalous emission lines that would be identified as technosignature candidates. However, the present configuration of PSG does not easily permit the inclusion of emission lines from artificial illumination or laser pulses, so the spectra shown in Figures 1 and 2 do not include these features. The estimates by T. G. Beatty (2022) have suggested that the scale of urbanization in S1 and S5 may be detectable by a large LUVOIR-like space telescope. Thus, for the present study, we will assume as a tentative hypothesis that HWO could conceivably detect emission features from nightside urban lighting in S1 and S5. However, further work is needed to assess how the magnitudes of the emission lines from large-scale urbanization (or laser pulses) would compare with other atmospheric spectral features when observed with HWO.

### 2.2. Observations of Other Planets

For an analog to the solar system, HWO would be able to detect not only Earth, but also other planets that can be observed in reflected light, such as Venus, Jupiter, and Saturn. Several scenarios involve technological activities extending from Earth to other parts of the solar system, including exploration, settlement, and even terraforming. Observations of other parts of the system can potentially provide additional or corroborating evidence of technosignatures in these scenarios. Several scenarios include settlements on the moons of the outer planets (S1, S3, S5, S9, and S10), but none of these would manifest as spectral signatures that could be observable in reflected-light observations of Jupiter or Saturn. Therefore, we will only focus on the possible reflected-light signatures that HWO could observe for future scenarios of Venus.

It is worth noting that the set of 10 scenarios includes several with significant technology on Mars, including industrial activities (S1, S2, S3, and S10), as well as full-scale planetary terraforming (S5 and S6). However, the reflected-light signature of Mars would be too faint to be observable by HWO, so none of the technosignatures associated with these activities on Mars would be detectable. Although human civilization today continues to build momentum toward the continued exploration and long-duration settlement of Mars (e.g., R. Zubrin & R. S. Wagner 1996; E. Musk 2017; J. Haqq-Misra 2023; C. Verseux et al. 2024), this expansion of Earth's technosphere to include Mars would not increase the visibility of such technospheres to remote observations. Likewise, the theoretical possibility that a planet like Mars might be utilized as a "service world" dedicated to industry (J. T. Wright et al. 2022) would not generate any spectral technosignatures that could be detected with HWO. A hypothetical planet larger than Mars might have a stronger reflected-light signal, but we refrain from considering any such planetary architectures in this study and instead focus only on solar system analogs.

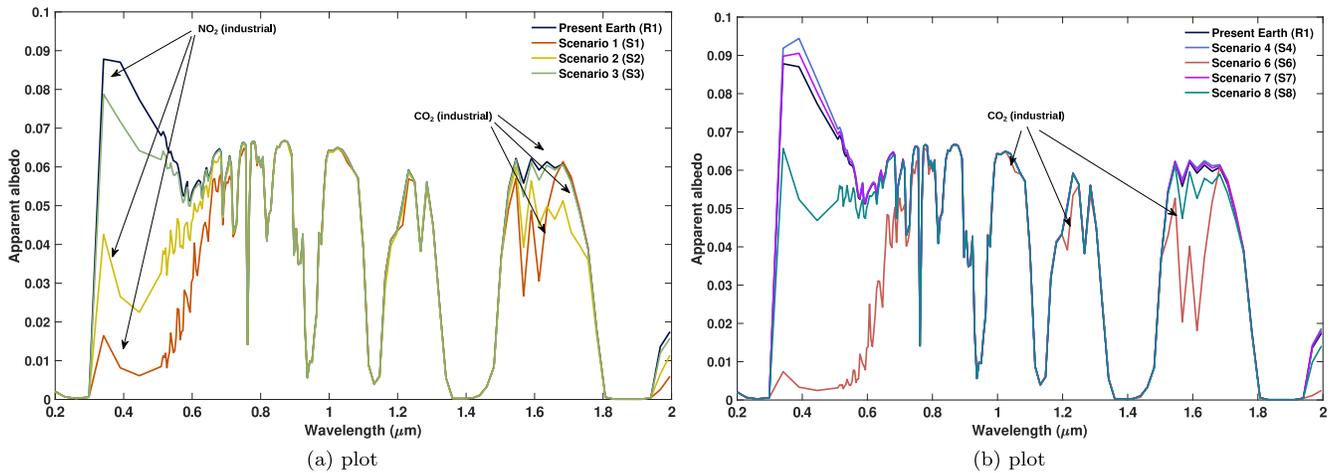

**Figure 2.** Albedo spectra of different scenarios from Table 1 compared to the present-day-Earth scenario (R1). Distinct absorption features from industrial activity—such as $NO_2$, in the short-wavelength region, and $CO_2$, in the near-infrared—are noticeable. Scenarios S5, S9, and S10 are omitted because they are spectrally identical to preagricultural Earth (R0; see Figure 1).

The atmospheric properties and compositions of future Venus for each of the 10 scenarios developed by J. Haqq-Misra et al. (2025a) are listed in Table 2. Six of these scenarios (S1, S2, S3, S4, S7, and S8) have atmospheres that are identical to present-day Venus, so these have been combined into a single column in Table 2. Terraforming on Venus occurs in scenarios S5 and S6. In both of these scenarios, large amounts of $CO_2$ have been removed to decrease the greenhouse effect, $SO_2$ abundances have been cut in half, and production of $O_2$ has begun (see M. Fogg 1995). Additional cooling occurs in S6 due to the use of starshades in orbit between the Sun and Venus, to decrease incident stellar radiation. Starshades are also used in S9 for even more aggressive cooling of the surface of Venus, but no other modification of the Venusian atmosphere occurs in S9. Scenario S10 involves large floating cities in the upper atmosphere of Venus, which support industrial activities and release pollutants into the Venusian atmosphere.

Venus may be technically observable with HWO, but none of the atmospheric constituents associated with technology in Table 2 would be detectable at the HWO wavelength range (see Figures 1 and 2). The sodium emission lines in S6 would be too dim to be observable with HWO, and the spectral signatures of industrial pollutants in S10 would be primarily at mid-infrared wavelengths and inaccessible to HWO (M. Elowitz 2022; J. Haqq-Misra et al. 2022b; S. Seager et al. 2023; E. W. Schwieterman et al. 2024). The primary differences between present-day Venus and any of the technologically modified future-Venus scenarios are due to the removal of $CO_2$ and $SO_2$ (S5 and S6) or the use of starshades to decrease surface temperature (S6 and S9). We do not perform any PSG calculations for these spectra, as the observation of a Venus-zone planet with depleted $CO_2$ and/or $SO_2$ would not necessarily constitute a technosignature. Although some of the scenarios include large-scale technological modifications of Venus, these changes would not be obvious with HWO observations.

**Table 2**
Atmospheric Properties for Future Venus in Each of the 10 Scenarios, with Reference Values for Present-day Venus (V0)

|  | V0 | S1–S4, S7, S8 | S5 | S6 | S9 | S10 |
|---|---|---|---|---|---|---|
| Mean Temperature (K) | 740 | 740 | 475 | 400 | 300 | 740 |
| Surface Pressure (bar) | 96 | 96 | 30 | 30 | 96 | 96 |
| $CO_2$ (%) | 96 | 96 | 96 | 96 | 96 | 96 |
| $N_2$ (%) | 3 | 3 | 3 | 3 | 3 | 3 |
| $SO_2$ (ppm) | 150 | 150 | 75 | 75 | 150 | 150 |
| $H_2O$ (ppm) | 50 | 50 | 50 | 50 | 50 | 50 |
| $O_2$ (ppm) | … | … | 300 | 300 | … | … |
| $NO_x$ (ppb) | … | … | … | … | … | 0.25 |
| CFC-11 (ppb) | … | … | … | … | … | 0.03 |
| CFC-12 (ppb) | … | … | … | … | … | 0.07 |
| $NF_3$ (ppt) | … | … | … | … | … | 0.33 |
| Na Emission (erg s$^{-1}$ cm$^{-2}$) | … | … | … | 0.001 | … | … |

**Note.** Six scenarios (S1, S2, S3, S4, S7, and S8) have atmospheric compositions identical to present-day Venus. One scenario includes high-pressure sodium emissions from inhabited settlements. Two scenarios (S5 and S6) involve ongoing terraforming to increase the habitability of Venus, by removing $CO_2$ and $SO_2$ while beginning to produce $O_2$. Two scenarios (S6 and S9) use starshades to cool the surface of Venus. One scenario (S10) involves industry in floating cities that release pollutants into the Venusian atmosphere.

### 3. Observations of Radio Emissions

Efforts to search for radio technosignatures from other stars have been ongoing since the 1960s, and current efforts by radio SETI groups around the world have begun to consider the prioritization of targets known or suspected to harbor exoplanets. One example was an observation campaign of over 2000 stars with exoplanet candidates or Kepler objects of interest, conducted at frequencies from 1 to 9 GHz, using the Allen Telescope Array (G. Harp et al. 2016). Another example was an observation of 61 TESS objects of interest at frequencies from 1 to 11 GHz using the Green Bank Telescope (N. Franz et al. 2022). Both of these search efforts could conceivably identify any narrowband radio transmissions that were directed toward Earth during the time of observations; the lack of any viable candidates from these search efforts implies that no persistent radio transmissions exist in this set of targets at the observed frequencies, assuming that the analysis pipeline would have been capable of recognizing such signals. Such radio technosignature searches cannot necessarily rule out periodic narrowband radio signals; they would only be able to do so by observing targets of interests all the time. Nevertheless, it remains likely that any potentially habitable planets that are discovered by missions like TESS and PLATO will continue to receive attention from such radio technosignature surveys. It likewise seems probable that any candidate habitable exoplanet systems that are observed by HWO—particularly if they show promise for hosting biosignatures—will be ideal targets to search for possible radio technosignatures.

The most obvious radio technosignatures in the scenarios developed by J. Haqq-Misra et al. (2025a) are the radio beacons in S1 and S5. Both of these scenarios include persistent and omnidirectional radio beacons that are constructed for the purpose of establishing communication with, and transmitting information to, life in other planetary systems (i.e., METI). Scenario S5 also includes a second intermittent METI transmission that involves targeted, rather than omnidirectional, broadcasts. We estimate the detectability of these radio beacons by assuming the transmitters are omnidirectional and continuous, broadcasting at a 10 GHz frequency (see J. Benford et al. 2010). The total integration time, $\tau_{\rm obs}$, that would be required to observe a radio beacon directed toward Earth can be obtained from the expressions given by S. Z. Sheikh et al. (2025):

$$\tau_{\rm obs} = \Delta\nu \left[ \frac{d^2 ({\rm SNR})({\rm SEFD})}{{\rm EIRP}_{\rm min}} \right]^2, \quad (2)$$

where SNR is the signal-to-noise ratio, SEFD is the system equivalent flux density, $\Delta\nu$ is the transmission bandwidth, and $\rm EIRP_{min}$ is a normalized transmitter power metric. Following the assumptions in the analysis by S. Z. Sheikh et al. (2025), we assume a receiver comparable to the Square Kilometre Array with $\rm SEFD = 1.5\,Jy$ and $\rm SNR = 5$. We show the dependence of the total integration time on the transmitter power in Figure 3, which includes a 10 Hz bandwidth case analogous to the Arecibo message and a 0.01 Hz bandwidth case comparable to radar science. We note that the maximum transmitter power of $10^{11}\,{\rm W}$ shown in Figure 3 is $\sim 0.1\%$ of the total energy use in S1 and S5 (see J. Haqq-Misra et al. 2025a, Table 9); such an expenditure would be a significant undertaking but could be a feasible investment for a civilization dedicated to a program of interstellar signaling. However, larger values of the transmitter power are probably unfeasible for these scenarios. A transmitter with $\rm EIRP_{min} = 10^{11}\,W$ would be detectable with the Square Kilometre Array at a total integration time of only 50 ms. At a lower transmitter power of $\sim 10^9\,{\rm W}$, comparable to the Deep Space Network, the integration time increases to 8.5 minutes. At even lower transmission powers, the required integration time becomes longer than the seconds to minutes that comprise typical radio

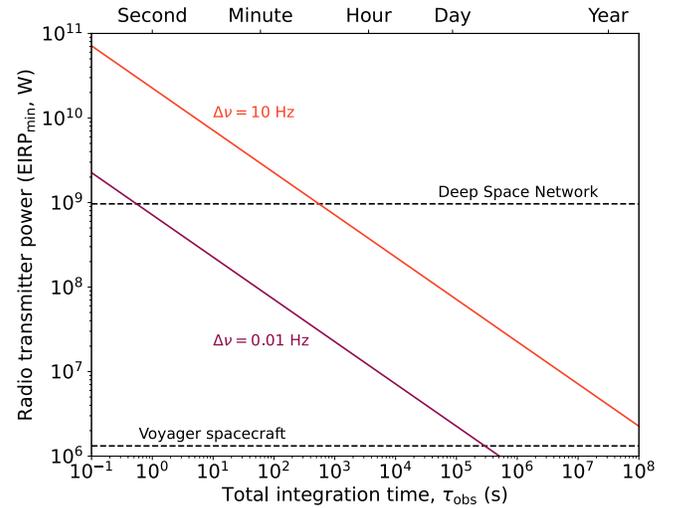

**Figure 3.** Total integration time required for the Square Kilometre Array to observe a radio beacon from scenarios S1 or S5 at a distance of 10 pc (following Equation (2)). The results are shown using a 10 Hz bandwidth analogous to the Arecibo message and a 0.01 Hz bandwidth comparable to radar science (see S. Z. Sheikh et al. 2025). The dashed lines show the comparable $\rm EIRP_{min}$ values for the Deep Space Network and Voyager spacecraft. The maximum transmission at $10^{11}$ W would correspond to $\sim 0.1\%$ of the scenario civilization's total power use, which could be detectable with 50 ms of observation time at 10 Hz. Lower-power transmissions, comparable to the Deep Space Network, would need 8.5 minutes, while transmitters comparable to Voyager would require over a year of integration time.

SETI observations, possibly requiring hours to days or longer to detect a signal, and even over a year for transmissions comparable to the Voyager spacecraft. These results indicate that even a dedicated beacon would not necessarily be easily detected with routine SETI observations; however, longer observations might be justifiable for planets that have already shown evidence of other potential biosignatures or technosignatures.

Observing radio technosignatures in the remaining eight scenarios is more difficult. All scenarios except S4 include radio communication between planets and from planets to satellites, but the likelihood of intercepting such signals during planet–planet occultations (e.g., N. Tusay et al. 2024; P. Fan et al. 2025) is very small. This paper assumes that the future scenarios all involve nontransiting systems, so any radio leakage would be unlikely to align with the observer and would not necessarily show obvious signs of periodicity. Likewise, several scenarios also include planetary defense radar signals (S1, S2, S3, S5, S6, S9, and S10), but the likelihood of intercepting such signals is also very small and thus not considered here.

## 4. Characterization by the LIFE Mission

The LIFE mission is a concept for a space-based and formation-flying nulling interferometer mission, which has been recommended for study by the European Space Agency (S. P. Quanz et al. 2022). The LIFE mission would enable the direct detection and atmospheric characterization of nearby exoplanets, including Earth-sized planets in the habitable zones of G dwarf host stars, at mid-infrared wavelengths ranging from $\sim 4$ to $18.5\,\mu{\rm m}$. Such observations would allow for atmospheric characterization that cannot be accomplished at the shorter-wavelength range of HWO. If the LIFE mission is approved and eventually constructed, then it would be logical for LIFE to conduct follow-up observations on any targets identified by HWO as promising for hosting biosignatures or technosignatures.

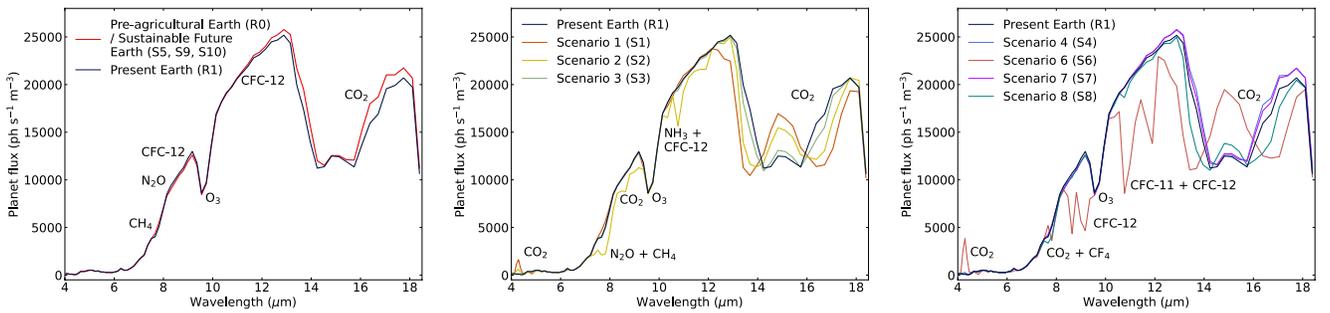

**Figure 4.** Mid-infrared spectra as observed by LIFE of preagricultural Earth (R0) and present-day Earth (R1) (left), scenarios S1, S2, and S3 (middle), and scenarios S4, S6, S7, and S8 (right). Scenarios S5, S9, and S10 are spectrally identical to R0. The calculations assume a 10 days integration period and a 10 pc distance to the planet. Some distinguishing features can be seen that result from differences in industry ($CO_2$, $NO_2$, CFC-11, and CFC-12) and large-scale agriculture ($N_2O$, $NH_3$, and $CH_4$).

The atmospheric compositions for future Earth in the scenario set (Table 1) include species that would be more detectable at mid-infrared wavelengths. For the future-Venus cases (Table 2), the abundances of such industrial constituents are probably too low to be detectable by the LIFE mission. Although the future scenarios involving a terraformed Mars (S5 and S6) would contain high abundances of such industrial greenhouse gases, the LIFE mission would not be able to observe any spectral features for a Mars-like planet (S. P. Quanz et al. 2022; Ó. Carrión-González et al. 2023). We use the LIFEsim software tool (F. A. Dannert et al. 2022) to simulate the LIFE observations for each scenario, shown in Figure 4. These LIFEsim calculations follow the configuration used in the study by E. W. Schwieterman et al. (2024), which includes a wavelength range from ∼4 to 18.5 μm, a spectral resolution of 50, a configuration with four 2 m apertures, and an assumption of three times the local zodiacal dust density. The target planet is assumed to reside 10 pc away, and the simulated observations use a 10 days integration period (see E. W. Schwieterman et al. 2024).

The left panel of Figure 4 shows the mid-infrared spectrum as observed by LIFE of preagricultural Earth (R0) and present-day Earth (R1). These two reference cases show some distinguishing features that result from differences in industrialization ($CO_2$ and CFC-12) and large-scale agriculture ($N_2O$ and $CH_4$); however, these differences are much less prominent than the visible-range $NO_2$ features shown in Figure 4. The mid-infrared spectra of the 10 scenarios as observed by LIFE are shown in the middle and right panels of Figure 4, indicating numerous distinct features of compounds that, if observed, would be evidence of a technosignature. Note that pronounced $CO_2$ features are clearly noticeable, particularly for the most heavily polluted scenarios (S1, S2, and S6). Although $CO_2$ itself would not be a technosignature, if it were to be observed coupled with the other industrial compounds, then this could indicate evidence of planetary-scale technological activity. In these spectra, CFC-11 and CFC-12 are the most dominant industrial compounds, potentially observable in scenarios S1, S2, S3, S6, and S8. Other pollutants from large-scale agriculture ($N_2O$, $CH_4$, and $NH_3$) are evident in the spectra of scenarios S2 and S8. Several scenarios include other industrial greenhouse gases (see Table 1), but in most cases these do not occur at high enough abundances to be obvious in the mid-infrared spectra (see E. W. Schwieterman et al. 2024). The exception is the combined $CO_2$ + $CF_4$ absorption feature at ∼7.8 μm in scenario S6, which would be a technosignature if observed. In general, the mid-infrared has more distinguishing features for identifying potential spectral technosignatures, and a mission concept such as LIFE may be able to identify these spectral technosignatures in any planets in the scenario set that involve significantly elevated abundances of such constituents.

## 5. Imaging with the Solar Gravitational Lens

A speculative but technically feasible approach to exoplanet observations is to use the Sun as a gravitational lens, with an optical telescope located at a large distance, in order to obtain resolved images of exoplanets. Proposals for developing such a solar gravitational lens would place the optical telescope at a distance of 650 to 1200 au from the Sun, which would allow for detailed observation of one or a few systems of interest along the line between the telescope and the Sun (e.g., C. Maccone 2009; S. G. Turyshev & V. T. Toth 2022; V. T. Toth & S. G. Turyshev 2023). The large-scale deployment of technology at continental scales, such as a planet-wide urban landscape, could conceivably be discerned from other biological phenomenon (S. Berdyugina & J. Kuhn 2019). The set of 10 scenarios developed by J. Haqq-Misra et al. (2025a) include several with large-scale surface features that may be identifiable with a solar-gravitational-lens mission. The use of a solar-gravitational-lens mission would be an intergenerational effort that would depend on the identification of specific targets that are worth investigating. If HWO were to observe a terrestrial planet in the habitable zone of its host star, perhaps with corroboration by a mission like LIFE, then this could be a stronger justification for constructing a solar-gravitational-lens mission to resolve images of this target system in an attempt to learn more about its surface or atmospheric features.

A solar gravitational lens would not have a single focal point but would instead have "a semi-infinite focal line"; this would render the image of an exoplanet as separated pixels within a cylindrical volume of diameter $2r_\oplus \simeq (\bar{z}/650\,\mathrm{au})$ $(30\,\mathrm{pc}/z_0)(1.34\,\mathrm{km})$, where $z_0$ is the distance to the exoplanet and $\bar{z}$ is the gravitational-lens baseline (S. G. Turyshev & V. T. Toth 2022). At a distance of $z_0 = 10$ pc, a solar-gravitational-lens mission would need to be located at $\bar{z} = 1030$ au, to potentially resolve continental-scale features on an Earth-sized planet (see, S. G. Turyshev & V. T. Toth 2022).

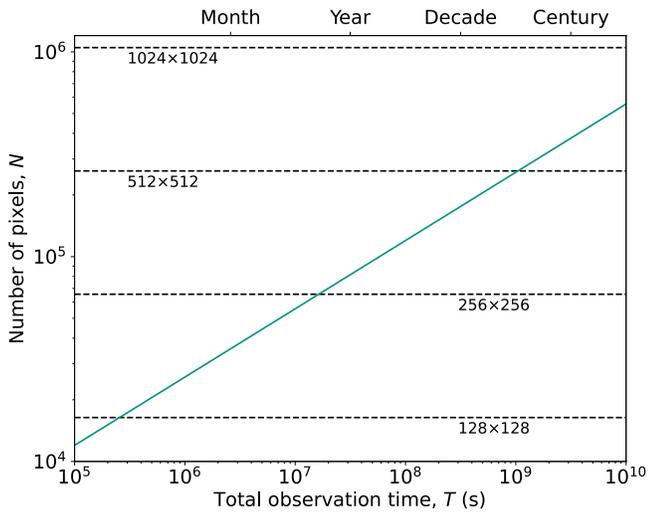

**Figure 5.** Total observation time required for a solar gravitational lens with a 1 m diameter optical telescope to resolve images of an Earth-sized planet at a distance of 10 pc (following Equation (3)). An image with $128 \times 128$ pixel resolution could be obtained in a few days, but higher resolutions of $256 \times 256$ would require half a year of observation time. Even higher resolutions would extend observation times to decades or even centuries.

The observation time, $T$, required to observe an exoplanet at a resolution of $N$ pixels was given by S. G. Turyshev & V. T. Toth (2022) as

$$T = N^2 \, \mathrm{SNR}^2 \left(\frac{1\,\mathrm{m}}{D}\right)^2 \left(\frac{z_0}{30\,\mathrm{pc}}\right)^2 \left(\frac{650\,\mathrm{au}}{\bar{z}}\right) \left(\frac{1\,\mu\mathrm{m}}{\lambda}\right) (1.21\,\mathrm{s}), \quad (3)$$

where $D = 2r_\oplus/\sqrt{N}$ is the distance between pixels, $\mathrm{SNR} = 5$, and the diameter of the optical telescope is 1 m. Figure 5 shows the observation time required to obtain a given pixel resolution for a gravitational-lens image of an Earth-like exoplanet at 10 pc. An image with a $128 \times 128$ pixel resolution could be resolved in 3 days, but increasing this resolution to $256 \times 256$ would require 0.5 yr total observation time. A resolution of $512 \times 512$ would require 33 yr, while a resolution of $1024 \times 1024$ pixels would be unattainable even with a century of observation time.

The most significant surface modifications for future Earth occur in S1 (with urban landscapes encompassing most of the continental land area on Earth), S5 (with nanoengineering on 90% of the planet's surface), and S6 (with nanoengineering on 80% of the planet's surface). For future Mars, both S5 and S6 include nanoengineering on 90% of the Martian surface, S9 includes technological networks across the entire Martian surface, and S10 includes built settlements across 88% of the Martian surface. For future Venus, S9 includes a dense cloud of technological network elements filling the entire Venusian atmosphere. Because such features span a majority of the planetary surfaces, a solar-gravitational-lens mission that resolves a $128 \times 128$ to $256 \times 256$ pixel image of any of these cases would be capable of providing evidence of exoplanetary surface technosignatures.

The surface features that occur in the remaining scenarios occupy much less of the total surface and may therefore be more difficult for a solar-gravitational-lens mission to identify as technological without higher-resolution images. About 6% of Earth's surface is modified in S2, 1% in S3, 0.1% in S7, and 3% in S8. Surface modifications on Earth are nearly negligible in S4 and nonexistent in S9 and S10. Scenarios S1, S2, and S3 include settlements on Mars, but these occupy too small a fraction of the surface to be resolvable. Scenario S6 includes 5% surface modification on Venus, as a result of ongoing terraforming activities, which may be difficult to resolve. Even a $1024 \times 1024$ pixel resolution, which would require longer than a century of observation time, would be insufficient to resolve such sparsely distributed surface features.

## 6. Discussion

The calculations in this paper of technosignature detectability with future missions are both encouraging and discouraging. It is encouraging that many of the technospheres in the scenarios developed by J. Haqq-Misra et al. (2025a) could be nominally detectable by HWO, but it is likewise discouraging that the most obvious signs of technology in many of these future projections of Earth's technosphere would remain undetected by future mission concepts. A summary of the ability of future missions to detect technosignatures in the set of 10 scenarios is provided in Figure 6. The most frequent technosignature that could be detectable by HWO in the scenario set is the $CO_2 + NO_2$ pairing, which arises from large-scale combustion and other industrial processes. The observation of both $CO_2$ and $NO_2$ increasing together may be interpreted as a tentative technosignature, although such a conclusion would need to rule out other false-positive explanations. Scenarios S1 and S5 also include large-scale nightside urban lighting that may be observed as sodium emission lines with HWO, which may be a more convincing technosignature, especially if observed in tandem with the $CO_2 + NO_2$ pairing, as in scenario S1. Follow-up radio observations would be able to corroborate the technospheres of S1 and S5 but would not identify any new ones. LIFE could provide evidence of scenarios with elevated abundances of artificial greenhouse gases and $CO_2$, which could reveal technospheres in scenarios S1, S2, S3, S6, and S8, depending on the observation time and mission architecture. A solar gravitational lens would corroborate S1, S5, and S6 and reveal technospheres in S9 and S10. Such an observing strategy involving HWO, radio, LIFE, and a solar-gravitational-lens mission would be able to find at least some candidate technosignatures in each of the 10 scenarios, although the cases with modestly elevated $CO_2$ and $NO_2$ but no other observables (S4 and S7) may be unconvincing as technosignature candidates.

The most conclusive way of identifying the presence of a technosphere in any of these scenarios would be to send a deep-space probe on a flyby mission (e.g., G. L. Matloff 2005) that is capable of obtaining high-resolution images of the exoplanetary surfaces (and perhaps even high-resolution spectroscopy of exoplanetary atmospheres). For scenarios S4 and S7, such follow-up observations with remote exploratory probes may be the only way to identify their technospheres. In all scenarios, large-scale alterations from urban landscapes or agriculture would be visible at resolutions of 50 m or greater; such features are present on Earth, Venus, and/or Mars in different scenarios (see Figure 6, last column). Missions such as the Lunar Reconnaissance Orbiter and the Mars Reconnaissance Orbiter have been able to capture images at resolutions of ~1 m or less; however, leveraging this technical capability for taking such images of exoplanets would require further advances in managing interstellar spaceflight and navigating a spacecraft close enough to an exoplanet. The only interstellar flyby mission that is currently under active development is Breakthrough Starshot, which would attempt to send a swarm of laser-propelled nanocraft toward the Alpha Centauri system (K. L. Parkin [2018](#)). Breakthrough Starshot may not necessarily be capable of the high-resolution imagery or spectroscopy that would be needed to reveal the technospheres in the set of 10 scenarios, but the Breakthrough Starshot model at least illustrates that the capability for interstellar exploration is conceptually attainable. It is somewhat sobering to consider the possibility that technospheres could be present on nearby exoplanetary systems, perhaps even abundant, but may be unrecognizable without actually sending spacecraft to visit them.

| SCENARIO | HABITABLE WORLDS OBSERVATORY | RADIO | LARGE INTERFEROMETER FOR EXOPLANETS | SOLAR GRAVITATIONAL LENS | DEEP SPACE PROBES |
|---|---|---|---|---|---|
| S1 | $CO_2$ + $NO_2$ Na Emission | radio beacon | $CO_2$ + CFC-11/12 | large surface features (Earth) | large surface features (Earth, Mars) |
| S2 | $CO_2$ + $NO_2$ | | $CO_2$ + CFC-11/12 $N_2O$ + $CH_4$ + $NH_3$ | | large surface features (Earth, Mars) |
| S3 | $CO_2$ + $NO_2$ | | $CO_2$ + CFC-11/12 | | large surface features (Earth, Mars) |
| S4 | $CO_2$ + $NO_2$ | | | | large surface features (Earth) |
| S5 | Na emission | radio beacons | | large surface features (Earth, Mars) | large surface features (Earth, Mars) |
| S6 | $CO_2$ + $NO_2$ | | $CO_2$ + CFC-11/12 $CF_4$ | large surface features (Earth, Mars) | large surface features (Venus, Earth, Mars) |
| S7 | $CO_2$ + $NO_2$ | | | | large surface features (Earth) |
| S8 | $CO_2$ + $NO_2$ | | $CO_2$ + CFC-11/12 $N_2O$ + $CH_4$ + $NH_3$ | | large surface features (Earth) |
| S9 | | | | large surface features (Venus, Mars) | large surface features (Venus, Mars) |
| S10 | | | | large surface features (Earth) | large surface features (Venus, Mars) |

**Figure 6.** Summary of technosignature detectability with future missions for each scenario. The darker colors indicate where multiple technosignatures in a scenario are observable by a mission.

It is important to note that the 10 scenarios analyzed in this study do not correspond to likelihoods or probabilities for technosignature detection. Each of the 10 scenarios represents a self-consistent projection of a future solar system technosphere that has a logical continuity from Earth today, and so in this sense they all represent *plausible* examples of technospheres that theoretically could exist. But this does not imply that all 10 scenarios have an equal weighting, nor does it imply that these 10 scenarios represent the complete set of possible technospheres that could exist. The methodological approach used by J. Haqq-Misra et al. ([2025a](#)) for developing these scenarios is sufficiently robust to have generated wide coverage of the possibility space given the assumptions underlying the futures projections, but it remains possible that relaxing or changing some of these assumptions could lead to other technospheres that are significantly different from those considered here. One assumption in these futures projections is that human civilization has not speciated, and another assumption is the projected timescale of 1000 yr into the future. Adjusting either of these assumptions to account for the future evolution of humans and other geologic-scale changes over greater timescales of 10,000 to 100,000 yr or longer could lead to different outcomes. Even more significantly, all of these scenarios begin with Earth today; however, it remains possible that any extraterrestrial technospheres that exist, and that could be detected, have little to no overlap with the technosphere on Earth today. If this is the case, then any reliance on Earth today as a guide for developing search strategies (whether through futures projections or other means) may be misguided. We cannot rule out such possibilities, but we also cannot easily "imagine the unimaginable." As a result, we are left with Earth as the only known example of technology, and any attempts to search for extraterrestrial technosignatures must inevitably at least consider the relevance of this single data point.

Another insight worth emphasizing is that all 10 of these scenarios involve situations of civilizations that tend to prioritize a strategy of exploration in spatial scale over exploitation in energy consumption. The analysis of these scenarios by J. Haqq-Misra et al. (2025b) showed that even in the scenario involving high rates of growth in energy consumption (1% annual growth for S9; see Table 9 of J. Haqq-Misra et al. 2025a), the expansion of the spatial domain (i.e., the rate at which the technosphere expands through the solar system) still proceeded more rapidly than increases in energy utilization (i.e., the rate at which the technosphere increases its energy demands). This is because the trajectories of all these scenarios fall well below the "luminosity limit," which is the maximum amount of energy available from harnessing luminous stellar energy within the spatial domain (J. Haqq-Misra et al. 2025b), analogous to the growth of an energy-intensive civilization as implied by applications of the Kardashev scale (see, e.g., M. Cirkovic 2015; R. H. Gray 2020). The fact that none of these scenarios involve rapidly expansive and energy-intensive growth raises questions about whether or not such concepts are ideal for guiding the search for technosignatures. The analysis by J. Haqq-Misra et al. (2025b) argues that the "luminosity limit" (or, equivalently, the limits implied by the Kardashev scale) may be unattainable, due to fundamental thermodynamic limits. It remains possible that some civilizational trajectories could opt to prioritize energy consumption (exploitation) over spatial growth (exploration), but such cases may have less continuity with present-day Earth. Exploring such possibilities remains a potentially viable option for developing alternative scenarios of detectable technospheres, although any such attempts will still inevitably be constrained by basic thermodynamic limits.

The idea of civilizational longevity is also worth revisiting. The 10 scenarios analyzed in this study all make projections of Earth's technosphere 1000 yr from now. This timescale is at least an order of magnitude larger than the duration of time that global-scale technology has existed on Earth, but it is still relatively small compared with the duration of life on Earth or other astrophysical phenomena. If very-long-lived technospheres exist (i.e, those that persist for at least millions, if not billions, of years), then these could conceivably be optimal targets for observation. However, it remains unknown if technological civilizations can persist for such long durations; for example, SETI pioneer Frank Drake speculated that the average detectable lifetime of technological civilizations is about 10,000 yr (F. Drake 2011). In this case, the 1000 yr future projections used in this study are a reasonable starting point for developing observation strategies, and future development of these scenarios could attempt extending the projections to 10,000 or 100,000 yr timescales, to obtain an even wider range of possibilities. Another consideration is the relative abundance of technospheres as compared to their duration: if short-lived technospheres are much more commonplace than long-lived ones, then those that are shorter-lived might be more easily observed (A. Balbi & C. Grimaldi 2024). Astrophysical observation of short-duration phenomena regularly occurs (examples include supernovae and gamma-ray bursts), so a short duration for the longevity of technospheres does not necessarily preclude their detectability. The analysis of the 10 scenarios in this study provides an example of the kinds of technosignatures that could conceivably be detectable, based on what is known about technology on Earth, and they can serve as motivation for continued thinking about the range of technospheres that may actually exist.

## 7. Conclusion

This study has presented 10 scenarios of Earth's 1000 yr future as reference cases for the detection of exoplanetary technospheres. Again, it is worth emphasizing at the conclusion that these scenarios all involve *projections* of possible futures, rather than *predictions* of a singular future, and so they should be interpreted accordingly as examples of the possibility space accessible to Earth today. In the search for technosignatures, this set of scenarios can be utilized as examples of exoplanet targets in order to determine the instrumental requirements, observation limits, and integration times that would be needed to detect specific technosignatures of interest.

A mission like the HWO would be able to reveal the presence of a $CO_2 + NO_2$ technosignature pair, resulting from combustion and other large-scale industry, in up to eight of the 10 scenarios. Follow-up observations with radio arrays could corroborate two of these technospheres through the discovery of a narrowband radio beacon. A mid-infrared mission such as LIFE could find further evidence of technospheres in five of these systems, including absorption features from $CO_2$, CFC-11, CFC-12, and $CF_4$, due to industry, as well as $N_2O$, $CH_4$, and $NH_3$, due to large-scale agriculture. Further observations with a solar-gravitational-lens mission could reveal large-scale surface features in five scenarios, including two that have not yet shown any technosignatures. Finally, the most conclusive way to search for technosignatures in all scenarios would be a flyby mission capable of resolving small-scale surface features on terrestrial exoplanets.

In all likelihood, no singular mission will be able to provide definitive evidence of an exoplanetary technosphere. Instead, corroborating evidence from different observatories, operating at different wavelengths, will be the most systematic and convincing approach toward identifying possible technosignatures. One important result from this study is that many atmospheric technosignatures in these scenarios come in pairs: $CO_2$ and $NO_2$ both increase together at visible/near-infrared wavelengths in seven scenarios, while mid-infrared technosignatures include $CO_2$ paired with other industrial gases in five scenarios, as well as an $N_2O + CH_4 + NH_3$ triple in two scenarios from agriculture. Further work should continue to explore other combinations of atmospheric constituents that, if observed together, would be evidence of extraterrestrial technology.

### Acknowledgments

Further details about the "Project Janus" scenarios of Earth's 1000 yr future are available at futures.bmsis.org. J.H.M. and R.K.K. gratefully acknowledge support from the NASA Exobiology program under grant 80NSSC22K1009. Any opinions, findings, and conclusions or recommendations expressed in this material are those of the authors and do not necessarily reflect the views of their employers or NASA.

*Software*: Planetary Spectrum Generator (G. L. Villanueva et al. 2018, 2022), LIFEsim (F. A. Dannert et al. 2022).

### ORCID iDs

Jacob Haqq-Misra https://orcid.org/0000-0003-4346-2611
Ravi K. Kopparapu https://orcid.org/0000-0002-5893-2471
George Profitiliotis https://orcid.org/0000-0002-4636-354X